\documentclass[conference]{IEEEtran}
\IEEEoverridecommandlockouts
\usepackage{cite}
\usepackage{amsmath,amssymb,amsfonts}
\usepackage{algorithm}
\usepackage{algpseudocode}
\usepackage{graphicx}
\usepackage{textcomp}
\usepackage{xcolor}
\usepackage{makecell}
\def\BibTeX{{\rm B\kern-.05em{\sc i\kern-.025em b}\kern-.08em
    T\kern-.1667em\lower.7ex\hbox{E}\kern-.125emX}}
\begin{document}

\title{Identifying Counterfeit Products using Blockchain Technology in Supply Chain System}

\author{\IEEEauthorblockN{Nafisa Anjum}
\IEEEauthorblockA{\textit{Department of Electronics and}\\{Telecommunication Engineering} \\
\textit{Chittagong University of}\\{Engineering and Technology}\\
Chittagong, Bangladesh \\
nafisaanjum94@gmail.com}
\and
\IEEEauthorblockN{Pramit Dutta}
\IEEEauthorblockA{\textit{Department of Electronics and}\\{Telecommunication Engineering} \\
\textit{Chittagong University of}\\{Engineering and Technology}\\
Chittagong, Bangladesh \\
pramitduttaanik@gmail.com}
}
\IEEEoverridecommandlockouts
\IEEEpubid{\makebox[\columnwidth]{978-1-6654-2678-7/22/\$31.00~\copyright2022 IEEE \hfill} \hspace{\columnsep}\makebox[\columnwidth]{ }}

\maketitle

\IEEEpubidadjcol

\begin{abstract}
With the advent of globalization and the ever-growing rate of technology, the volume of production as well as ease of procuring counterfeit goods has become unprecedented. Be it food, drug or luxury items, all kinds of industrial manufacturers and distributors are now seeking greater transparency in supply chain operations with a  view to deter counterfeiting. This paper introduces a decentralized Blockchain based application system (DApp) with a view to identifying counterfeit products in the supply chain system. With the rapid rise of Blockchain technology, it has become known that data recorded within Blockchain is immutable and secure. Hence, the proposed project here uses this concept to handle the transfer of ownership of products. A consumer can verify the product distribution and ownership information scanning a Quick Response (QR) code generated by the DApp for each product linked to the Blockchain.
\end{abstract}

\section{Introduction}
Over the years, the identification of counterfeit goods in market has always posed a challenge for all supply chain stakeholders. As per the latest assessment of EU Intellectual Property Office (EUIPO) and the Organization for Economic Cooperation and Development (OECD), the global sales of counterfeit and pirated goods have increased alarmingly to 460 billion euros which is about 3.3\% of the global trade [1]. The sales and profits of companies around the world have been affected by this phenomenon. The clothing and pharmaceutical sectors experienced sales losses of about 26.3 billion euros and 10.2 billion euros respectively [2]. Moreover, with the advent of recent technologies and E-commerce, the market of counterfeit goods have exploded on social media platforms. The anonymity, reach and segmenting tools of Ecommerce and social media have smoothed the pavement for counterfeiters. Hence, counterfeiting as in producing twins or fakes of real products pose great threat to innovation and economic growth.\newline

Blockchain technology has been receiving much attention over the past decade and its numerous applications are being developed. Blockchain is a decentralized system of shared, immutable ledger. It facilitates the process of recording, trading and tracking assets over a business network thus reducing risks and cutting costs for all involved. Hence, any application using Blockchain as its base technology ensures that the data are tamper resistant.\newline

In this paper, a decentralized application system (DApp) has been introduced that uses Ethereum blockchain technology in its architecture. The DApp simulates a real world supply chain and ensures the ownership of product is transferred and recorded in the blockchain network. Besides, the system proposed here can also be implemented in Ecommerce and retail sites that can considerably bring transparency in the virtual platforms for all consumers. Though Radio Frequency Identification (RFID) has been used for research in this sector previously, it has posed security and privacy risks which can be efficiently dealt with using blockchain.

\begin{figure*}[htbp]
\centering
\includegraphics[width=\textwidth,height=0.35\textheight]{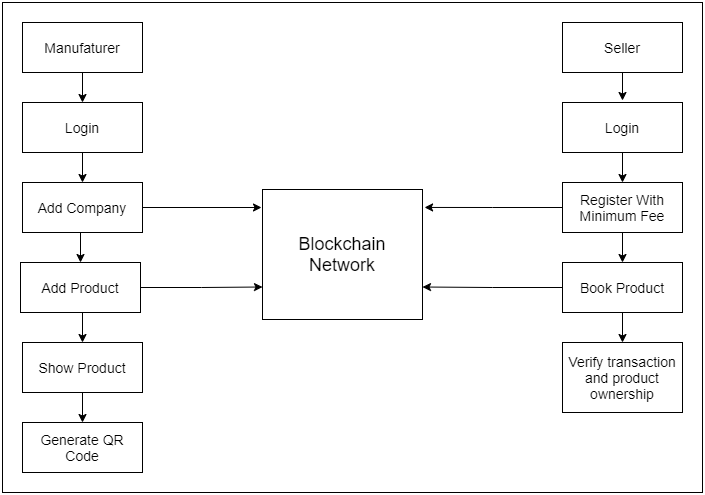}
\caption{System Diagram}
\end{figure*}

\section{Related Work}

Various researchers have proposed different methods for establishing a blockchain-based supply chain management system. One of them presented a counterfeit product identification system using android application where a product can be searched in the Blockchain network [3]. Another paper displayed a fake product detection system using blockchain where SHA-256 Algorithm was used to identify a product [4]. A fully functional anti-product forgery system was designed by a group of researchers that uses digital signature for verification [5]. In one of the papers, a blockchain-based Product Ownership Management System was proposed. It displayed the use of blockchain-based system over traditional RFID-based system [6]. Another paper presented a food traceability system using IoT and blockchain collectively. In this model, they used fuzzy logic to evaluate the food quality [7]. A paper displayed a system where blockchain was used with RFID to remove the limitation in post supply chain [8]. To improve the current supply chain method a paper used blockchain combined with IoT to track product origin [9].

\section{Methodology}
The system proposed here uses MetaMask cryptocurrency wallet for transactions and and the smart contract here has been deployed in the Rinkeby Test Network of the Ethereum Blockchain. The DApp is based on three major stakeholders, the Manufacturer, the Seller and the Consumer.
\subsection{System Diagram}
Fig. 1 depicts the system diagram of the proposed DApp. Every user of the DApp has to be authenticated before logging in. This authentication system has been implemented using Firebase which is a platform provided by Google for developing interactive mobile and web applications. After successful authentication, the manufacturer can add their company to the DApp and enroll products of the company. The contract address of the company is provided to the manufacturer and all the company data as well as manufacturer’s account address are stored in the blockchain network. After a product has been included in the blockchain, it is assigned a QR code for verification. The sellers can buy products from manufacturer after registration. The ownership transfer of the product can be tracked through the QR code.

\subsection{Manufacturer}
The manufacturer’s functions include adding the company to the blockchain by providing company name and setting the minimum registration fee to become a seller or retailer for the company. The manufacturer solely preserves the rights to enroll products in the network. The manufacturer can also control the distribution status of products and transfer ownership after a seller has bought the product stock.

\begin{figure}[htbp]
\centering
\includegraphics[width=0.4\textwidth,height=0.23\textheight]{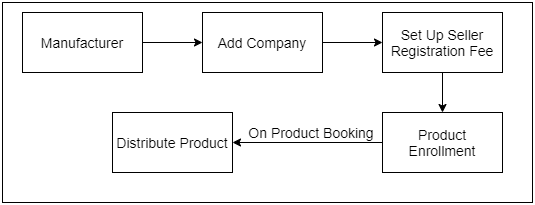}
\caption{Manufacturer's Working Process}
\label{fig}
\end{figure}

The manufacturer performs two major functions namely adding and distributing products in this system. For adding a product Algorithm 1 is used.

\begin{algorithm}
\caption{Create Product}\label{alg:cap}
 \hspace*{\algorithmicindent} \textbf{Input:} Product Name, Product Price, Product Stock \\
 \hspace*{\algorithmicindent} \textbf{Output:} Added Product 
\begin{algorithmic}
\If{$msg.sender$ is not manufacturer}
    \State throw;
    \State end
\Else
    \State insert product in product array
\EndIf
\end{algorithmic}
\end{algorithm}

For distribution of product Algorithm 2 is used. The product and order status in the blockchain is changed through this.

\begin{algorithm}
\caption{Distribute Product}\label{alg:cap}
 \hspace*{\algorithmicindent} \textbf{Input:} Product ID \\
 \hspace*{\algorithmicindent} \textbf{Output:} Changed Product Status
\begin{algorithmic}

\If{$msg.sender$ is not manufacturer}
    \State throw;
    \State end
\Else
    \State change product status to 'Shipped' and set order status as complete
\EndIf
\end{algorithmic}
\end{algorithm}

\subsection{Seller}
A seller can pay the minimum fee set by the manufacturer and register for the company. After registering once, the seller can buy any product as well as track its distribution. A product status is set from ‘Ready To Go’ to ‘Shipped’ after the manufacturer ships it out to the seller.
\
\begin{figure}[htbp]
\centering
\includegraphics[width=0.4\textwidth,height=0.25\textheight]{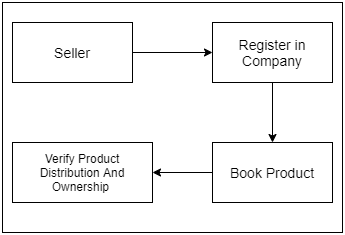}
\caption{Seller's Working Process}
\label{fig}
\end{figure}

Algorithm 3 here is used to make sure a seller pays the minimum registration fee set by the manufacturer.

\begin{algorithm}
\caption{Seller Registration}\label{alg:cap}
 \hspace*{\algorithmicindent} \textbf{Input:} Minimum amount of registration fee set by manufacturer \\
 \hspace*{\algorithmicindent} \textbf{Output:} Registered Seller
\begin{algorithmic}

\If{$msg.sender$ is registered seller or fee is less than requirement}
    \State throw;
    \State end
\Else
    \State map $msg.sender$ is true
\EndIf
\end{algorithmic}
\end{algorithm}

Algorithm 4 here is used by the seller to buy or book products from the manufacturer. It records the seller's data in the blockchain.

\begin{algorithm}
\caption{Buy Product}\label{alg:cap}
\hspace*{\algorithmicindent} \textbf{Input:} Product ID, Seller Name, Amount to buy\\
\hspace*{\algorithmicindent} \textbf{Output:} Set Current Owner of product as $msg.Sender$
\begin{algorithmic}

\If{$msg.sender$ is not registered seller}
    \State throw;
    \State end
\ElsIf{$msg.value$ is less that required amount}
    \State throw;
    \State end
\Else
    \State set product owner name as seller name and store account address of seller
\EndIf
\end{algorithmic}
\end{algorithm}

\subsection{Consumer}
A consumer can scan the QR code provided with each product and verify the transfer of ownership of product from manufacturer to seller. The consumer can also verify the name of the current owner of the product and check its distribution status.

\begin{figure*}[htbp]
\centering
\includegraphics[width=\textwidth,height=0.4\textheight]{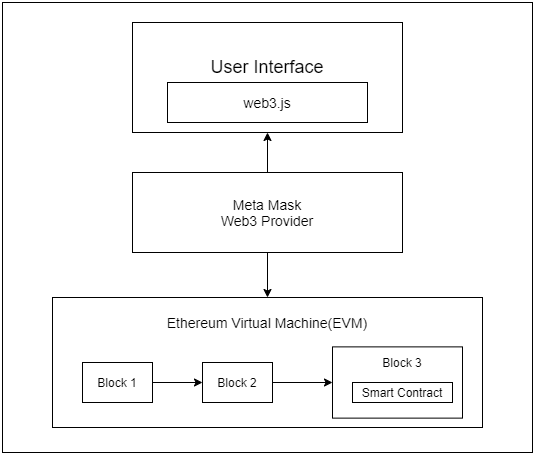}
\caption{Ethereum DApp Architecture}
\end{figure*}

\subsection{Blockchain}
Blockchain technology provides promising opportunities in the supply chain management paradigm. Blockchain data is stored on nodes where each node has a complete copy of the blockchain database. Orders, payments, accounts, price of products etc. can be tracked, shared and secured using a blockchain network. Some important features of blockchain technology in supply chain management includes:
\subsubsection{Security and Privacy}
Blockchain uses public key encryption method of cryptography for data security. Users have public and private key pair which are used to validate transactions and these transactions are immutable and permanent.
\subsubsection{Decentralization}
As blockchain is a distributed ledger technology, it doesn’t rely on third party or any centralized authority.
\subsubsection{Transparency}
Data stored in Blockchain is public and anyone can enquire on their transactions. The transactions can be governed by a set of rules known as the smart contract. \\
The system proposed here uses MetaMask cryptocurrency wallet for transactions and and the smart contract here has been deployed in the Rinkeby Test Network of the Ethereum Blockchain. The DApp is based on three major stakeholders, The Manufacturer, the seller and the consumer.

\subsection{Ethereum DApp Architecture}
Fig. 4 here depicts the base architecture of the system. The user interface (UI) here has been developed using ReactJS. If the user wants to interact with the smart contract, the DApp will use Web3.js which communicates with MetaMask through its provider. MetaMask creates a transaction and signs it with the user’s private key. This transaction is then sent to Ethereum network. The transaction is processed, verified and added to a block in the network. The private keys of the user are never recorded in the process so user can safely interact with the network.

\section{Result Analysis}
Sending data to the Blockchain comes with some cost referred to as transaction cost. Miners tend to prioritize transactions with higher costs. Transaction cost is measured in gas and gas fees are paid in Ethereum’s native currency ether (ETH). The table indicates the transaction cost and gas fees required for the proposed system.

\begin{table}[H]
\caption{Cost Calculation}
\begin{center}
\begin{tabular}{|c|c|c|c|}
\hline
\bfseries\makecell{Sl No.}&\bfseries\makecell{Function\\ Description}&\bfseries\makecell{Transaction\\ Cost (gas)}&\bfseries\makecell{Gas\\ Fee (ETH)}  \\
\hline
1&\makecell{Deploy Contract\\ of our system}&133405&0.001333 \\
\hline
2&\makecell{Adding\\ New Company}&1068597&0.001069 \\
\hline
3&\makecell{Seller Registration}&45755&0.000046 \\
\hline
4&\makecell{Product Enrollment}&208571&0.000209 \\
\hline
5&\makecell{Buying Product}&41581&0.000042 \\
\hline
6&\makecell{Product Distribution}&55578&0.000056 \\
\hline
\multicolumn{4}{|c|}{Total=0.002755 ETH/ \$8.56} \\
\hline
\multicolumn{4}{|c|}{Deploy= \$4.14} \\
\hline
\end{tabular}
\label{tab1}
\end{center}
\end{table}

Here CoinMarketCap [10] was used to convert Ether to US dollars.  Remix which is a web browser IDE for developing DApp was used determine the gas needs. MetaMask was used for contract interaction and determining the costings. The cost for deploying our contract in the Rinkeby Test Network is 0.001333 ETH which is equivalent to 4.14 US dollars. The overall costing for the system is less than 10 US dollars which proves the cost effectiveness of the proposed model. The product ownership transfer as well product quality assurance costs are also reduced here compared to current market trends to verify product authenticity.
A consumer can scan the QR code and verify the ownership transfer of the product. The manufacturer’s account address, the seller’s account address and name as well as the status of product is recorded in the QR code. If the product status is ‘Shipped’, the product transfer is genuine and the order is set to ‘complete’ in the blockchain. The QR code is provided with copy-sensitive digital image pattern.

\section{Conclusion}
Ownership tracking system is being reshaped through distributed ledgers of Blockchain technology. Due to rapid changes in the Ecommerce and business sectors, the current trends of supply chain are being affected. The DApp developed here ensures greater transparency in the supply chain management and can also be entrusted for use in Ecommerce. As such, administrative costs and complicated procedures are eliminated by this process. Besides, the cost for enrolling each product in the proposed model is only 0.000209 ether which is equivalent to 0.65 US dollars that can sufficiently reduce costs for large chain stores. The model also ensures end-user verification system through a QR code and transactions here can be verified on Etherscan too. As future work of the proposed model, the functions included can be improved further to bring reliability in the supply chain management.

\end{document}